\documentclass[sigconf]{acmart}
\AtBeginDocument{%
  \providecommand\BibTeX{{%
    \normalfont B\kern-0.5em{\scshape i\kern-0.25em b}\kern-0.8em\TeX}}}

\setcopyright{acmcopyright}
\copyrightyear{2022}
\acmYear{2022}

\acmConference[MM '22]{Proceedings of the 30th ACM International Conference on Multimedia}{October 10--14, 2022}{Lisboa, Portugal}
\acmBooktitle{Proceedings of the 30th ACM International Conference on Multimedia (MM '22), October 10--14, 2022, Lisboa, Portugal}
\acmPrice{15.00}
\acmDOI{10.1145/3503161.3548068}
\acmISBN{978-1-4503-9203-7/22/10}

\usepackage{bbding}

\acmSubmissionID{1308}

\begin{document}

\title{Model-Guided Multi-Contrast Deep Unfolding Network 
for MRI Super-resolution Reconstruction}

%

\author{Gang Yang}
\email{yg1997@mail.ustc.edu.cn}
\orcid{0000-0001-9403-5818}
\affiliation{%
  \institution{University of Science and Technology of China}
  \city{}
  \state{}
  \country{}
  \postcode{43017-6221}
}

\author{Li Zhang}
\email{zanly20@mail.ustc.edu.cn}
\orcid{0000-0003-1610-6056}
\affiliation{%
  \institution{University of Science and Technology of China}
  \city{}
  \state{}
  \country{}}
  
\author{Man Zhou}
\email{manman@mail.ustc.edu.cn}
\orcid{0000-0003-2872-605X}
\affiliation{%
  \institution{University of Science and Technology of China}
  \city{}
  \state{}
  \country{}}

\author{Aiping Liu}
\authornote{Corresponding author: Aiping Liu}
\email{aipingl@ustc.edu.cn}
\orcid{0000-0001-8849-5228}
\affiliation{%
  \institution{University of Science and Technology of China}
  \institution{USTC IAT-Huami Joint Laboratory for Brain-Machine Intelligence, Institute of Advanced Technology}
  \city{}
  \country{}}

\author{Xun Chen}
\email{xunchen@ustc.edu.cn}
\orcid{0000-0002-4922-8116}
\affiliation{%
  \institution{University of Science and Technology of China}
  \institution{USTC IAT-Huami Joint Laboratory for Brain-Machine Intelligence, Institute of Advanced Technology}
  \city{}
  \country{}}

\author{Zhiwei Xiong}
\email{zwxiong@ustc.edu.cn}
\orcid{0000-0002-9787-7460}
\affiliation{%
  \institution{University of Science and Technology of China}
  \institution{Institute of Artificial Intelligence, Hefei Comprehensive National Science Center}
  \city{}
  \country{}
}

\author{Feng Wu}
\email{fengwu@ustc.edu.cn}
\affiliation{%
  \institution{University of Science and Technology of China}
  \institution{Institute of Artificial Intelligence, Hefei Comprehensive National Science Center}
  \city{}
  \country{}
}

%

\renewcommand{\shortauthors}{Gang Yang et al.}
\renewcommand{\authors}{Gang Yang, Li Zhang, Man Zhou, Aiping Liu, Xun Chen, Zhiwei Xiong, and Feng Wu}

%
\begin{abstract}
Magnetic resonance imaging (MRI) with high resolution (HR) provides more detailed information for accurate diagnosis and quantitative image analysis.
Despite the significant advances, most existing super-resolution (SR) reconstruction network for medical images has two flaws:
1) All of them are designed in a black-box principle, thus lacking sufficient interpretability and further limiting their practical applications. Interpretable neural network models are of significant interest since they enhance the trustworthiness required in clinical practice when dealing with medical images.
2) most existing SR reconstruction approaches only use a single contrast or use a simple multi-contrast fusion mechanism, neglecting the complex relationships between different contrasts that are critical for SR improvement.
  To deal with these issues, in this paper, a novel \textbf{M}odel-\textbf{G}uided interpretable \textbf{D}eep \textbf{U}nfolding \textbf{N}etwork (\textbf{MGDUN}) for medical image SR reconstruction is proposed. 
  The Model-Guided image SR reconstruction approach solves manually designed objective functions to reconstruct HR MRI. We show how to unfold an iterative MGDUN algorithm into a novel model-guided deep unfolding network by taking the MRI observation matrix and explicit multi-contrast relationship matrix into account during the end-to-end optimization. 
  Extensive experiments on the multi-contrast IXI dataset and BraTs 2019 dataset demonstrate the superiority of our proposed model.
  
\end{abstract}



\maketitle

\section{Introduction}

Magnetic resonance imaging (MRI) has been widely adopted in clinical and medical research. In comparison to other imaging modalities such as computed tomography (CT) and nuclear imaging, MRI enjoys the advantage of delivering detailed images of tissue architecture without the use of ionizing radiation~\cite{feng2021MINet}.
The MRI system may be configured in a number of ways using pulse sequences to provide multi-contrast images such as T1, T2, and proton density (PD) weighted images that include essential physiological and pathological features. 
However, in real-world cases, HR MR images are often obtained with a longer scanning time, lower signal-to-noise ratio, and small spatial converge~\cite{zhang2021mr, plenge2012super}. 
Additionally, the quality of MR images acquired in clinical practice may be insufficient owing to patients' involuntary physiological movements (e.g., heart pounding and breathing) during the acquisition process.
This is particularly problematic when protocols requiring a long echo time (TE) or repetition time (TR) are used. 
These scans may lead to inaccurate diagnosis as limited structural and textural information is provided in subsequent quantitative medical image analysis ~\cite{feng2021exploring}. 
As a result, there is emerging interest in developing super-resolution (SR) techniques for reconstructing high-resolution (HR) outputs from low-resolution (LR) images to increase the spatial resolution of magnetic resonance imaging.

The utility of super-resolution is capable to improve the quality of MR images without modifying the hardware and overcome the challenges in obtaining HR MRI scans.
  MRI super-resolution approaches can be broadly classified into two categories depending on the number of imaging modalities involved: single-contrast super-resolution (SCSR) methods and multi-contrast super-resolution (MCSR) methods.
  SCSR approaches have been extensively studied over the last several decades, with the goal of reconstructing the high-resolution counterpart of a given low-resolution image in a single contrast mode, thereby ignoring the complementary multi-contrast information.
  In contrast to SCSR methods, MCSR methods recover a target modality by synthesizing information from multiple modalities. Clinically, MRI generates multi-contrast images under a variety of imaging settings but with the same anatomical structure, which includes T1 and T2 weighted images (T1WIs and T2WIs), as well as proton density and fast-suppressed proton density weighted images (PDWIs and FS-PDWIs), providing complementary information to each other~\cite{zeng2018simultaneous, mai2011robust}.
  Noting that contrasts with shorter acquisition times are easier to obtain, they can be used to supplement a single LR image with extra information. For example, relevant HR information from T1WIs or PDWIs may be utilized as auxiliary contrasts to aid in the generation of target contrasts.

  Existing techniques for image super-resolution reconstruction include model-based and learning-based approaches. Model-based techniques utilize domain knowledge when modeling the physical mechanism underlying the issue. Typical optimization algorithms include alternating direction method of multipliers (ADMM) algorithm~\cite{sun2016deep}, and iterative shrinkage-thresholding algorithm (ISTA)~\cite{zhang2018ista}. Regardless of its theoretical attractiveness, model-based approaches are incapable of performing end-to-end optimization, resulting in limited performance. 
  Alternatively, deep learning-based SR approaches have gained growing attention in recent years. For example, various architectures such as residual networks~\cite{chaudhari2018super}, generative adversarial networks~\cite{lyu2020multi}, and densely connected networks~\cite{chen2018brain} are utilized to reconstruct an MR image. Nevertheless, neural networks lack transparency (i.e., the black-box design) with generalized structures, and it is unclear how domain knowledge can be incorporated. 
  When dealing with medical images, the accuracy and trustworthiness of reconstruction are critical for discovery and diagnosis. 
  Therefore, balancing accuracy and interpretability is a non-trivial problem. The objective of designing interpretable neural networks is to bridge the gap between model-based and learning-based methods which can be accomplished by unfolding the iterations of an inference algorithm into deep neural networks, thus making the learning process interpretable.
  
  In this paper, a novel \textbf{M}odel-\textbf{G}uided interpretable \textbf{D}eep \textbf{U}nfolding \textbf{N}etwork (\textbf{MGDUN}) for medical image SR reconstruction is proposed. The motivation of our approach has two folds. On the one hand, to fully exploit domain knowledge of MRI SR and improve prediction performance, we formulate two manually designed objective functions for reconstructing HR MRI, each corresponding to a recovery process and incorporating domain knowledge.
  We then show how to solve these functions iteratively and how to unfold the iterative MGDUN algorithm into a neural network form by implementing specially designed modules.
  In contrast to conventional neural networks, model-guided design results in transparent network architectures that are well-aligned with the emerging interpretable machine learning framework.
  On the other hand, we formulate MGDUN as a network to reconstruct an HR image of a target contrast from an LR input with the aid of other guide contrasts, which is capable to exploit the complex relationships between different modalities better.  
  
  The main contributions of this paper are as follows:
  \begin{itemize}
      \item We design a novel \textbf{M}odel-\textbf{G}uided \textbf{D}eep \textbf{U}nfolding \textbf{N}etwork (\textbf{MGDUN}) for medical image SR reconstruction, which models multi-contrast MRI SR with other contrast images in an interpretable manner.
      \item We elaborate on how to solve the manually designed objective functions and how to unfold the iterative algorithm into a neural network by incorporating domain knowledge with specially designed modules.
      \item We reconstruct an HR image of a target contrast from an LR input with the aid of other guide contrasts, providing a new strategy for multi-contrast fusion.
	 \item Extensive experiments on the multi-contrast IXI dataset and BraTs 2019 dataset demonstrate the superiority of MGDUN.
  \end{itemize}

\section{RELATED WORK}

\subsection{Multi-contrast MR image representation}

Clinically, MR images are usually acquired with multiple contrasts under a variety of imaging settings for comprehensive evaluation~\cite{lyu2020multi,feng2021multi}, and each provides unique and complementary structural information about tissues~\cite{zeng2018simultaneous, brown2011mri}. 
As a result, Multi-contrast has been proposed to improve representation ability for a variety of MR image tasks, including segmentation and SR. 
For example, Huo et al. trained and evaluated MRI segmentation using T1WIs and T2WIs.~\cite{huo2018splenomegaly}. 
Different from the multi-contrast MRI segmentation task, the SR task requires the division of images into auxiliary and target contrasts. The auxiliary contrast, which is usually easier to obtain, is utilized to assist the reconstruction of the target contrast. Rough et al., for instance, super-resolved the target image using anatomical intermodality priors from a reference image~\cite{rousseau2010non}. Meanwhile, similar anatomical structures in the auxiliary contrast may be utilized to reconstruct an SR image from its LR counterpart~\cite{jafari2014mri, zheng2017multi, zheng2018multi}. However, most current methods are limited in constructing a model-based interpretable network, making the investigation of the relationship between multiple contrasts challenging.

\subsection{Medical Image Super-Resolution}

Medical image super-resolution methods are classified into two broad categories: single-contrast super-resolution (SCSR)~\cite{lim2017enhanced, zhao2018deep, KUKLISOVAMURGASOVA20121550, pham2017brain, mcdonagh2017context, zhao2019channel, zhang2018image, feng2021T2Net} and multi-contrast super-resolution (MCSR) ~\cite{feng2021exploring, zeng2018simultaneous, zheng2017multi, lu2015mr, manjon2010mri}.
Traditionally, because of their simplicity, bicubic and b-spline interpolations are two of the most frequently used SCSR methods in MRI practice.
However, both methods invariably generate fuzzy edges and block artifacts. To address these issues, EDSR and MDSR~\cite{lim2017enhanced} employ multiple blocks with linear residuals, contributing to improved performance in image super-resolution, whereas RCAN~\cite{zhang2018image} employs numerous residual groups with long skip connections and several residual blocks with short skip connections within each residual group. Recently, T$^2$Net~\cite{feng2021T2Net} with joint MRI reconstruction and SR enables representation and feature sharing across tasks, producing higher-quality, super-resolved, and motion-artifact-free images.
In another line of studies, MCSR methods have shown superior performance and the ability to make full use of domain knowledge from multiple modalities. For example, Feng et al.~\cite{feng2021exploring} develop a novel and successful solution called SANet, which consists of a separable attention network for exploring foreground and background areas in forward and backward directions using auxiliary contrast.

\subsection{Deep unfolding network}
As a pioneer work, deep unfolding is first reported in ~\cite{gregor2010learning}, and it designs a learned version of the iterative soft thresholding algorithm (ISTA) that can be unfolded into a neural network form. Since then, a series of works~\cite{kokkinos2018deep,you2021ista, song2021memory,ning2020accurate,zhang2020deep} demonstrate that deep unfolding methods are applicable to certain optimization algorithms since they can not only optimize the parameters in an end-to-end manner by minimizing the loss function over a large training set, but also integrate model-based and learning-based methods well. For instance, a novel fast network~\cite{afonso2010fast,ning2020accurate} based on half-quadratic splitting is proposed to solve the unconstrained optimization problem in the task of image restoration and reconstruction. 
Another important work in image SR ~\cite{zhang2020deep} proposes an end-to-end trainable unfolding network that integrates the flexibility of model-based methods with the advantages of learning-based methods. Additionally, CoISTA~\cite{deng2019deep} introduces a novel joint multi-modal dictionary learning (JMDL) method for modeling cross-modal dependency. It converts the JMDL model into a deep neural network by unfolding the iterative shrinkage and thresholding algorithm (ISTA).

\section{METHOD}

 \begin{figure*}[ht]
   \centering
   \includegraphics[width=\linewidth]{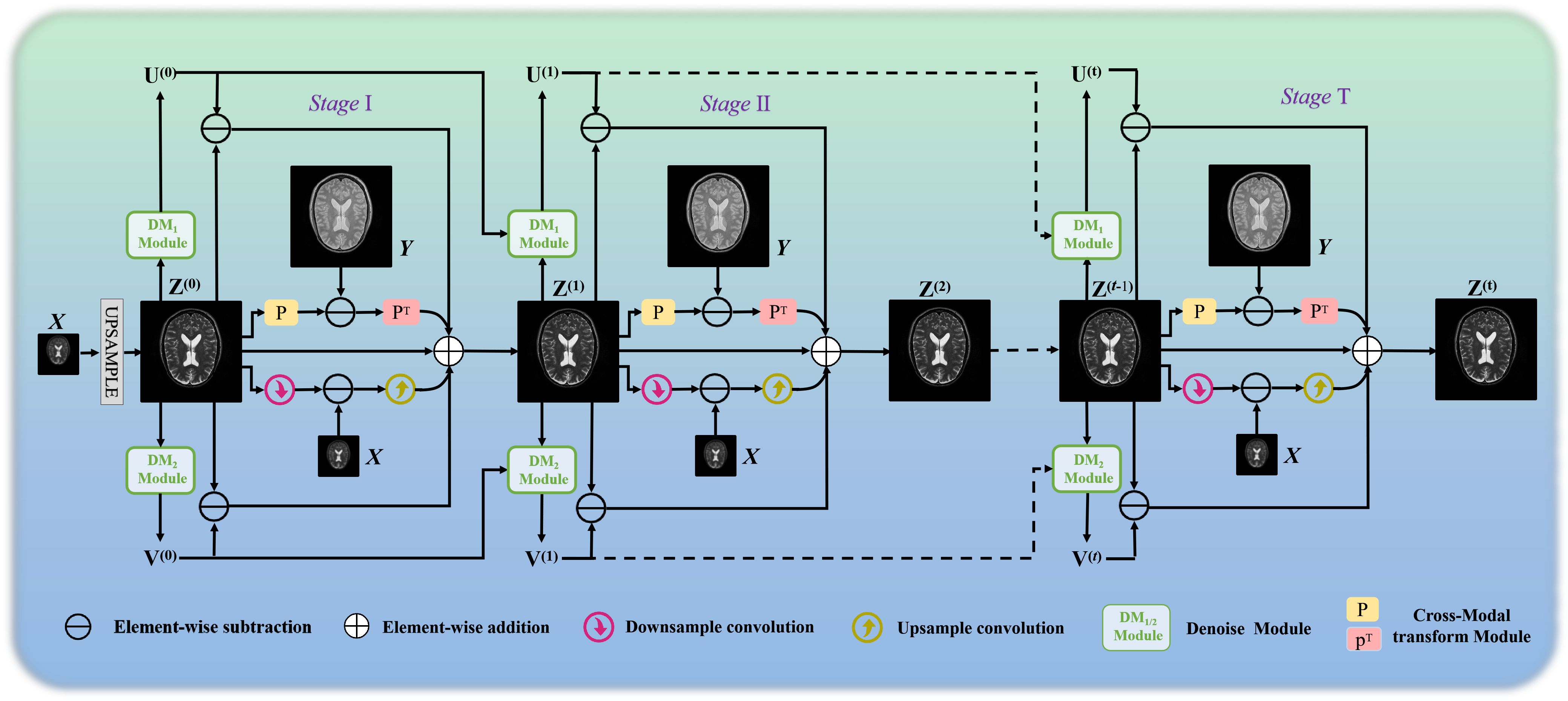}
   \caption{The overall architecture of MGDUN. It is a model-guided interpretable network with T stages.}
   \label{fig:overall}
 \end{figure*}

\subsection{Motivation}
An HR image $I_{HR}$ can return the LR image $I_{LR}$ obtained following the down-sampling process, and the process of down-sampling $f$ can be expressed as follows:
\begin{align}
    I_{LR}\;=\;f(I_{HR})\;=\;\phi(I_{HR})\;+\;{\mathcal N}
\end{align}
where $\phi$ denotes the function for down-sampling or blurring, and ${\mathcal N}$ denotes the system noise. The SR process is theoretically aimed at exploring the inverse solution $f^{-1}$ to the original down-sampling function $f$. Owing to the fact that the SR process is an ill-posed problem, it is impossible to obtain an exact inverse solution; only approximate solutions are possible. The goal of the SR imaging process is to find the most desirable inverse function $g$ of the theory inverse solution $f^{-1}$. 
\begin{align}
    I_{SR}\;=\;g(I_{LR})\;\approx\;I_{HR}
\end{align}
where $I_{SR}$ denotes the corresponding SR image. To obtain such an approximate solution $g$, it is necessary to use the image prior. There are still limitations on the prior information of single contrast images, so we take advantage of multi-contrast MR images. Based on the prior information of multi-contrast MRI, we propose a novel \textbf{M}odel-\textbf{G}uided interpretable \textbf{D}eep \textbf{U}nfolding \textbf{N}etwork (\textbf{MGDUN}) for medical image SR reconstruction reconstruction.

\subsection{Model-Guided MRI SR algorithm}  \label{sec:MGDUN_algorithm}
\subsubsection{The objective functions}
\ 
\newline 
Let $X \in \mathbb{R}^{n\times C}$ represents the degraded observations and $Z \in \mathbb{R}^{N\times C}$ represents the unknown original image, where $C$ denotes the number of channels, and $n=h \times w$ and $N=H \times W$. It is assumed that an LR image is obtained by down-sampling and blurring an HR image, so the linear relationship between the observed image and the original HR image can be typically formulated as follows:
\begin{align}
    X\;=\;DKZ\;+\;{\mathcal N}_1
\end{align}
where $D \in \mathbb{R}^{n \times N}$ and $K \in \mathbb{R}^{N \times N}$ represent the process of down-sampling and blurring, respectively, and ${\mathcal N}_1$ denotes the noise.

Transform  modal relationship considering multi-contrast MR images in MCSR task, the  transform relationship between the guide image $Y \in \mathbb{R}^{N\times C}$ and the unknown original image can be formulated by:
\begin{align}
    Y\;=\;PZ\;+\;{\mathcal N}_2
\end{align}
where $P \in \mathbb{R}^{N \times N}$ is the transform function, and the ${\mathcal N}_2$ represents the noise in this process.
As a result, $Z$ can be obtained by solving the following objective function:
\begin{align}
    Z\;=\;\underset Z{argmin}&\frac12{\vert\vert X\;-\;DKZ\vert\vert}_2^2\;+\;\frac\eta2{\vert\vert Y\;-\;PZ\vert\vert}_2^2\;   \notag \\
    &+\;\lambda_1{\mathfrak R}_1(Z)\;+\;\lambda_2{\mathfrak R}_2(Z)   \label{eq:objective_function}
\end{align}
where the hyper parameters ($\eta, \lambda_1, \lambda_2$) are the trade-off coefficient, and the last two regularization terms correspond to the prior domain knowledge of the MCSR task, which are noise prior in the typical degradation process and transform modal noise prior in multi-contrast image, respectively. The choice of various regularization functions reflects different ways of incorporating prior knowledge about the unknown original HR MR image.

In the next section, we describe how to solve the objective function with an iterative algorithm. Then, we unfold the iterative algorithm into neural networks for image SR, obtaining an end-to-end reconstruction architecture.

\subsubsection{The Proximal Gradient Descent Algorithm}
\ 
\newline

Following the framework of half-quadratic splitting (HQS) to introduce two auxiliary splitting parameters $U$ and $V$ for $Z$ with different prior knowledge of the MCSR task, the Eq.~\ref{eq:objective_function} can be formulated as a non-constrained optimization problem, which can be written as:
\begin{align}
    \underset{Z,\;U,\;V}{argmin}&\frac12{\vert\vert X\;-\;DKZ\vert\vert}_2^2\;+\;\frac\eta2{\vert\vert Y\;-\;PZ\vert\vert}_2^2\;  \notag \\
    &+\frac{\beta_1}2{\vert\vert U\;-\;Z\vert\vert}_2^2\;+\frac{\beta_2}2{\vert\vert V\;-\;Z\vert\vert}_2^2\;  \notag \\
    &+\;\lambda_1{\mathfrak R}_1(U)\;+\;\lambda_2{\mathfrak R}_2(V)     \label{eq:HQS}
\end{align}
where $\beta_1$, $\beta_2$, $\lambda_1$, and $\lambda_2$ are the penalty parameters. 
To obtain an unrolling inference, Eq.~\ref{eq:HQS} can be divided into the following three sub-problems and solved alternatively:
\begin{align}
    U^{(t)}\;=\;&\underset U{argmin}\frac{\beta_1}2{\vert\vert U\;-\;Z^{(t)}\vert\vert}_2^2\;+\;\lambda_1{\mathfrak R}_1(U)\;  \\
    V^{(t)}\;=\;&\underset V{argmin}\frac{\beta_2}2{\vert\vert V\;-\;Z^{(t)}\vert\vert}_2^2\;+\;\lambda_2{\mathfrak R}_2(V)\;  \\
    Z^{(t+1)}\;=\;&\underset Z{argmin}\frac12{\vert\vert X\;-\;DKZ\vert\vert}_2^2\;+\;\frac\eta2{\vert\vert Y\;-\;PZ\vert\vert}_2^2\;    \notag\\
    &+\frac{\beta_1}2{\vert\vert U^{(t)}\;-\;Z\vert\vert}_2^2\;+\frac{\beta_2}2{\vert\vert V^{(t)}\;-\;Z\vert\vert}_2^2\;
\end{align}
here, $t$ denotes the HQS iteration index.

For the objective function of Eq.~\ref{eq:objective_function} with the prior knowledge of the MCSR task, we employ the efficient Proximal Gradient Descent (PGD) to solve the above three sub-problems:
\begin{align}
    U^{(t)}\;&=\;{Prox}_{{\mathfrak R}_1}(U^{(t-1)}\;-\;\delta_1\nabla_U\;\mathcal F(U^{(t-1)}))\;  \notag  \label{eq:iter_U}  \\
    &=\;\;{Prox}_{{\mathfrak R}_1}(U^{(t-1)}\;-\;\delta_1(\beta_1(U^{(t-1)}\;-\;Z^{(t)}))) \\
    V^{(t)}\;&=\;{Prox}_{{\mathfrak R}_2}(V^{(t-1)}\;-\;\delta_2\nabla_V\;\mathcal F(V^{(t-1)}))\; \notag \label{eq:iter_V} \\
    &=\;\;{Prox}_{{\mathfrak R}_2}(V^{(t-1)}\;-\;\delta_2(\beta_2(V^{(t-1)}\;-\;Z^{(t)})))       \\
    Z^{(t+1)}\;&=\;Z^{(t)}\;-\;\delta_3\nabla_Z\;\mathcal F(Z^{(t)})   \label{eq:iter_Z}
\end{align}
where $Prox_{{\mathfrak R}_1}(\cdot)$ and $Prox_{{\mathfrak R}_2}(\cdot)$ are proximal operators corresponding to penalty ${{\mathfrak R}_1}(\cdot)$ and ${{\mathfrak R}_2}(\cdot)$, which integrate the information coming from target contrast LR MRI and guidance contrast HR MRI. And the gradient related notations are detailed as:
\begin{align}
    \nabla_Z\;\mathcal F(Z^{(t)})\;=\;&{(DK)}^T{(DKZ^{(t)}\;-\;X})\;+\;\eta P^T(PZ^{(t)}\;-\;Y)\; \notag \\ &+\beta_1(Z^{(t)}\;-\;U^{(t)})+{\beta_2}(Z^{(t)}\;-\;V^{(t)})   \label{eq:fz}
\end{align}

In summary, the iterative algorithm for solving the MCSR task of Eq.~\ref{eq:objective_function} is given above, where we initialize $Z^{(0)}$ with a bicubic interpolated version of $X$. Under the framework of the interpretable MGDUN model, the PGD algorithm usually requires dozens of iterations to converge.

 \begin{figure*}[ht]
   \centering
   \includegraphics[width=0.82\linewidth]{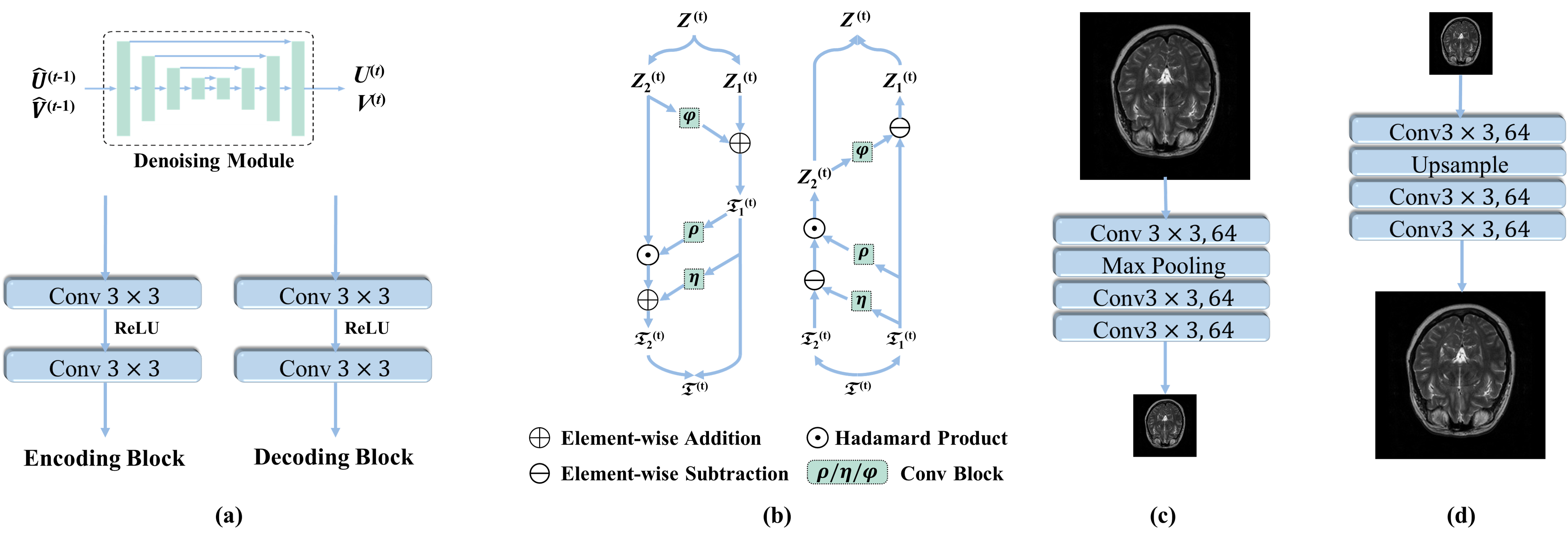}
   \caption{Architectures of MGDUN's Submodules. (a) Denoising Module; (b) INN blocks in cross-modal transform module; (c)  Up-sampling blocks ($\boldsymbol {Up}$); (d) down-sampling blocks ($\boldsymbol{Down}$).}
   \label{fig:submodule}
 \end{figure*}

\subsection{Model-Guided deep unfolding network}

In Sec.~\ref{sec:MGDUN_algorithm}, we have proposed a general model-guided MRI SR algorithm for MCSR tasks. Like other model-based image restoration, it is difficult to optimize Eq.~\ref{eq:iter_U} and Eq.~\ref{eq:iter_V} due to the nonlinearity. Meantime, in traditional model-based approaches~\cite{boyd2011distributed, he2016deep}, alternatively solving the above three optimization problems requires many iterations to converge leading to prohibitive computational cost. An alternative approach is to unfold the iterative optimization into a series of network implementations as demonstrated in recent years~\cite{wisdom2017building, chen2018deep, dong2018denoising, bertocchi2020deep}. The total number of unfolding stages naturally corresponds to that of PGD iterations.

\subsubsection{Model represent and model overview}
\ 
\newline

The main idea behind deep unfolding network is that conventional iterative soft-thresholding algorithm (ISTA) can be implemented equivalently by a stack of recurrent neural networks~\cite{wisdom2017building}. Inspired by the principle of model-driven deep learning, we generalize Eq.~\ref{eq:iter_U} to Eq.~\ref{eq:iter_Z} as a network block. Each step is translated with deep learning terminologies.

In each network, two auxiliary variables ($U$ and $V$) are updated first. Due to the existence of noise, the $Prox(\cdot)$ operator can be implemented by a deep denoising module ($\boldsymbol{DM}$). In Eq.~\ref{eq:iter_U}, given the evaluated HR image $Z^{(t)}$ of the current stage and auxiliary splitting parameter $U^{(t-1)}$ of the previous stage, it generates the auxiliary splitting parameter $U^{(t)}$ of the current stage. The same as $V^{(t)}$ in the form of Eq.~\ref{eq:iter_V}. In neural networks, these two steps are implemented by:
\begin{align}
    U^{(t)}\;=&\;\boldsymbol{{DM}_1}(U^{(t-1)}\;+\;\xi_1Z^{(t)}; C, C', C)      \\
    V^{(t)}\;=&\;\boldsymbol{{DM}_2}(V^{(t-1)}\;+\;\xi_2Z^{(t)}; C, C', C)      
\end{align}
where $\boldsymbol{DM}(\cdot; C, C', C)$ is used to obtain more expressive auxiliary splitting parameter. $C'$ is the number of channels for feature maps.

Then, we reconstruct our estimated HR image according to Eq.~\ref{eq:iter_Z} and Eq.~\ref{eq:fz}. In our network, the reconstruction process which is implemented by a reconstruction module consists of cross-modal transform module, down-sampling block, and up-sampling block, takes auxiliary various $U^{(t)}$, $V^{(t)}$, the evaluated HR image $Z^{(t)}$, the LR MRI and the guide image $Y$ as inputs and outputs the reconstructed image $Z^{(t+1)}$.
Therefore, the reconstruction process in the neural network is as follows:
\begin{align}
    Z^{(t+1)}\;=&\;Z^{(t)}\;-\;\;\delta_3(\boldsymbol {Up}(\boldsymbol {Down}(Z^{(t)})\;-\;X)\;   \notag \\
    &+\;\eta\boldsymbol P^{\boldsymbol T}(\boldsymbol P\boldsymbol( Z^{(\mathbf t)}\boldsymbol)\;-\;Y)\;  \notag \\ 
    &+\beta_1(Z^{(t)}\;-\;U^{(t)})+{\beta_2}(Z^{(t)}\;-\;V^{(t)})\;)\;  \label{eq:reconstruction}
\end{align}
where $\boldsymbol {Up}(\cdot)$ and $\boldsymbol{Down}(\cdot)$ denote the up-sampling and down-sampling operator in spatial resolution, respectively, and $\boldsymbol P(\cdot)$ and $\boldsymbol {P^T}(\cdot)$ perform the cross-modal transform functions.

Then, the updated $Z^{(t+1)}$ is fed into the next stage to refine the estimate $U$ and $V$ again. The denoising module and the reconstruction module are alternatively updated $T$ times until reaching the final reconstruction.

The overall network architecture of the MGDUN is shown in Fig.~\ref{fig:overall}, which contains $T$ stages that are intentionally designed to correspond to $T$ iterations in the PGD optimization algorithm. Each stage of MGDUN consists of three specified network modules, containing a deep denoising module for denoising and updating for auxiliary variables, a cross-modal transform Module for cross-modal transform function, and a reconstruction module for reconstruction and updating for $Z$.
We will elaborate on each module next. 

\subsubsection{The deep denoising module}
\vspace{1em}
\ 
\newline

The design of deep denoising modules corresponds to the computing the updated estimate $U^{(t)}$ and $V^{(t)}$ in Eq.~\ref{eq:iter_U} and Eq.~\ref{eq:iter_V}, respectively. In general, any existing image denoising network can be used as the denoising module here.

As shown in Fig.~\ref{fig:overall},  the intermediate estimates $U^{(t-1)}$ and $V^{(t-1)}$ are fed into the proximal operator after weighting with the intermediate estimate $Z^{(t)}$ for further refinement. In this paper, we have adopted a variant of U-net as the backbone of the deep denoising module. Other more effective networks for medical image denoising can be also adopted.
The U-net denoising network consists of an encoder and a decoder.
As shown in Fig.~\ref{fig:submodule}(a), the encoder consists of four encoding blocks, which contain two convolutional layers with $3\times 3$ kernels and ReLU nonlinearity. Corresponding to the encoder, the decoder also consists of four decoding blocks, which contain two convolutional layers with $3\times 3$ kernels and ReLU nonlinearity. Instead of predicting the refine auxiliary various directly, we enable the denoising module to predict the residual by adding a skip connection from the input to the output.
To reduce the number of network parameters and the effect of overfitting, we opt to enforce all denoising modules sharing the same network parameters.

\subsubsection{Cross-Modal Transform Module}
\vspace{0.5em}
\ 
\newline

Noted that Eq.~\ref{eq:reconstruction} involves the cross-modal transform matrix $\boldsymbol P$ and $\boldsymbol {P^T}$ that are expensive to calculate. 
We find that $\boldsymbol P$ and $\boldsymbol {P^T}$ perform a cross-modal transform operation, and the two processes are inverted: one from the target contrast image to the guide contrast image, and the other from the guide contrast image to the target contrast image. The process can be formulated as:
\begin{align}
    \mathfrak T^{(t)}\;=&\;\boldsymbol P(Z^{(t)})        \label{eq:P}\\
    Z^{(t)}\;=&\;\boldsymbol P^{\mathbf T}(\mathfrak T^{(t)})    \label{eq:pt}
\end{align}
As a result, we design cross-modal transform modules using the principle of invertible neural networks (INNs).
INNs have been adopted in various inference tasks and achieved excellent performance due to their flexibility~\cite{kingma2016improved, lu2021large}.
We formulate an INN architecture design to serve as cross-modal transform operation. It consists of two pixel shuffling layers~\cite{dinh2016density} and several INN blocks.
As shown in Figure \ref{fig:submodule}(b), relevant invertible modules are embedded in the cross-modal transform module. 

For the t-th stage, given an evaluation HR MRI ($Z^{(t)}$) to be refined, we first put it to one pixel shuffling layer for changing dimension, then pass through several INN blocks to execute the cross-modal transform function, and finally restore the original dimension through the other pixel shuffling layer.

For the forward operation, one pixel shuffling layer executes dimension addition first. 
Then the input ($Z^{(t)}$) is divided into ($Z_1^{(t)}$) and ($Z_2^{(t)}$) along the channel axis, and the corresponding cross-modal transform output is $\mathfrak T_1^{(t)}$ and $\mathfrak T_2^{(t)}$ (two components of $\mathfrak T^{(t)}$). This process corresponds to the operation of cross-modal transform matrix $\boldsymbol P$, in which an INN block can be expressed as :
\begin{align}
    \left\{\begin{array}{l}\mathfrak T_1^{(t)}\;=\;Z_1^{(t)}\;+\;\varphi(Z_2^{(t)})\\\mathfrak T_2^{(t)}\;=\;Z_2^{(t)}\odot exp(\rho(\mathfrak T_1^{(t)}))\;+\;\eta(\mathfrak T_1^{(t)})\end{array}\right.	\label{eq:inn_P}
\end{align}
where $\varphi(\cdot)$, $\eta(\cdot)$ and $\rho(\cdot)$ are arbitrary functions, $exp(\cdot)$ is Exponential functions, and $\odot$ is the Hadamard product.

Accordingly, for the backward operation, given [$\mathfrak T_1^{(t)}$, $\mathfrak T_2^{(t)}$], it is easy to calculate [$Z_1^{(t)}$, $Z_2^{(t)}$] as:
\begin{align}
    \left\{\begin{array}{l}Z_2^{(t)}\;=\;(\mathfrak T_2^{(t)}\;-\;\eta(\mathfrak T_1^{(t)}))\odot exp(-\rho(\mathfrak T_1^{(t)}))\\Z_1^{(t)}\;=\;\mathfrak T_1^{(t)}\;-\;\varphi(Z_2^{(t)})\end{array}\right.		\label{eq:inn_PT}
\end{align}
This process corresponds to the operation of matrix $\boldsymbol {P^T}$ in Eq.~\ref{eq:pt}. The forward and backward operations are shown in Fig.~\ref{fig:submodule}.

\subsubsection{Reconstruction Module}
\ 
\newline

The design of the reconstruction module corresponds to the update of intermediate evaluated $Z^{(T)}$ as described in Eq.~\ref{eq:reconstruction}.
With the output of denoising modules ($U^{(t)}$, $V^{(t)}$), the evaluated HR image $Z^{(t)}$, the LR MRI and the guide image $Y$, we can reconstruct the updated image $Z^{(t+1)}$. The architecture of the reconstruction module is shown in Fig.~\ref{fig:overall} and Eq.~\ref{eq:reconstruction}, which still involve the degradation operations (${(DK)^T}$ and ${DK}$). The pair of operation ${(DK)^T}$ and ${DK}$ can be implemented by up-sampling and down-sampling layers for modeling capability.

The operators ${(DK)^T}$ and ${DK}$ are simulated using a convolution network layer respectively. Specifically, ${DK}$ is simulated by a network called down-sampling-blocks ($Down$) consisting of a convolutional layer with $3\times3$ kernels and 64 channels, one max pool layer to decrease the spatial resolution, and two convolutional layers with $3\times3$ kernels for reprojection to the original dimension (as shown in Figure~\ref{fig:submodule}(c)).
Similarly,  the ${(DK)^T}$ is simulated by a network call Up-sampling-blocks ($Up$) consisting of a convolutional layer with $3\times3$ kernels and 64 channels, and one upsample layer to increase the spatial resolution and two convolutional layers with $3\times3$ kernels for reprojection to the original dimension as shown in Figure~\ref{fig:submodule}(d).

\subsubsection{network training}
\ 
\newline

We will apply this model for the super-resolution reconstruction of LR T2WI with the aid of HR PDWI (or HR T1WI). 

Our MGDUN is supervised by the $\mathcal{L}_1$ loss between $Z^{(T)}$ and the groundtruth $Z$. Then, the overall network is trained by minimizing the following loss function:
\begin{align}
    \Theta\;=\;\underset\Theta{argmin}\sum_{i=1}^N{\left\|g(X_i,\;Y_i;\;\Theta)\;-\;Z_i\right\|}_1
\end{align}
where $X_i$, $Y_i$, and $Z_i$ denote the $i^{th}$ pair of target contrast LR T2WI, guide HR PDWI, and the original target contrast HR T2WI, respectively. $g(\cdot;\Theta)$ denotes the reconstructed HR T2WI by the network with parameter $\Theta$.

\begin{table*}[]
\centering
\caption{The comparisons of average PSNR, SSIM, and MSE on IXI and BraTs datasets with $2\times$ and $4\times$ enlargements. The best and second best results are highlighted in \textcolor{red}{\textbf{red}} and \textcolor{blue}{\underline{blue}} color, respectively. The up or down arrows indicate higher or lower values corresponding to better results.}
\label{tab:main_metrics}
\resizebox{\textwidth}{!}{%
\begin{tabular}{c|cccccc|cccccc}
\hline
Datasets    & \multicolumn{6}{c|}{IXI}                                                     & \multicolumn{6}{c}{BraTs}                                                    \\ \hline
Scales      & \multicolumn{3}{c|}{2$\times$ SR}                       & \multicolumn{3}{c|}{4$\times$ SR}  & \multicolumn{3}{c|}{2$\times$ SR}                       & \multicolumn{3}{c}{4$\times$ SR}   \\ \hline
Metrics     & PSNR $\uparrow$   & SSIM $\uparrow$   & \multicolumn{1}{c|}{MSE $\downarrow$}     & PSNR $\uparrow$   & SSIM $\uparrow$  & MSE $\downarrow$    & PSNR $\uparrow$   & SSIM $\uparrow$  & \multicolumn{1}{c|}{MSE $\downarrow$}     & PSNR $\uparrow$    & SSIM $\uparrow$  & MSE $\downarrow$    \\ \hline
Bicubic     & 24.5537 & 0.7584 & \multicolumn{1}{c|}{15.4654} & 24.3658 & 0.7508 & 15.8045 & 26.5771 & 0.8413 & \multicolumn{1}{c|}{12.1369} & 26.3637 & 0.8360 & 12.4472 \\
EDSR~\cite{lim2017enhanced}        & 31.4066 & 0.9290 & \multicolumn{1}{c|}{7.0586}  & 29.5377 & 0.9028 & 8.7625  & 33.2810 & 0.9557 & \multicolumn{1}{c|}{5.6549}  & 31.9166 & 0.9419 & 6.6225  \\
MDSR~\cite{lim2017enhanced}        & 30.9519 & 0.9242 & \multicolumn{1}{c|}{7.5883}  & 29.6663 & 0.9042 & 8.6254  & 33.3624 & 0.9567 & \multicolumn{1}{c|}{5.5991}  & 32.0175 & 0.9430 & 6.5372  \\
RCAN~\cite{zhang2018image}        & 31.9391 & 0.9351 & \multicolumn{1}{c|}{6.6395}  & 31.3783 & 0.9301 & 7.0721  & 34.1138 & 0.9631 & \multicolumn{1}{c|}{5.1123}  & 32.7279 & 0.9504 & 6.0093  \\
CoISTA~\cite{deng2019deep}      & 31.4199 & 0.9121 & \multicolumn{1}{c|}{7.0657}  & 29.4435 & 0.8757 & 8.516   & 29.2042 & 0.9028 & \multicolumn{1}{c|}{9.088}   & 27.4678 & 0.8812 & 11.0865 \\
T$^2$Net~\cite{feng2021T2Net}       & 30.0556 & 0.9117 & \multicolumn{1}{c|}{8.2676}  & 29.4629 & 0.9014 & 8.8596  & 32.9922 & 0.9530 & \multicolumn{1}{c|}{5.8623}  & 31.7000 & 0.9389 & 6.8210  \\
SANet~\cite{feng2021exploring}       & \textcolor{blue}{\underline{36.7565}} & \textcolor{blue}{\underline{0.9683}} & \multicolumn{1}{c|}{\textcolor{blue}{\underline{3.7998}}}  & \textcolor{blue}{\underline{35.2765}} & \textcolor{blue}{\underline{0.9616}} & \textcolor{blue}{\underline{4.5115}}  & \textcolor{blue}{\underline{35.0775}} & \textcolor{blue}{\underline{0.9636}} & \multicolumn{1}{c|}{\textcolor{blue}{\underline{4.4661}}}  & \textcolor{blue}{\underline{33.5158}} & \textcolor{blue}{\underline{0.9578}} & \textcolor{blue}{\underline{4.9808}}  \\ \hline
MGDUN(Ours) & \textcolor{red}{\textbf{37.3366}} & \textcolor{red}{\textbf{0.9691}} & \multicolumn{1}{c|}{\textcolor{red}{\textbf{3.5598}}}  & \textcolor{red}{\textbf{35.9786}} & \textcolor{red}{\textbf{0.9637}} & \textcolor{red}{\textbf{4.1639}}  & \textcolor{red}{\textbf{35.9690}} & \textcolor{red}{\textbf{0.9703}} & \multicolumn{1}{c|}{\textcolor{red}{\textbf{4.1529}}}  & \textcolor{red}{\textbf{34.5577}} & \textcolor{red}{\textbf{0.9615}} & \textcolor{red}{\textbf{4.8933}}  \\ \hline
\end{tabular}%
}
\end{table*}

\begin{figure*}[t]
  \centering
  \includegraphics[width=0.81\linewidth]{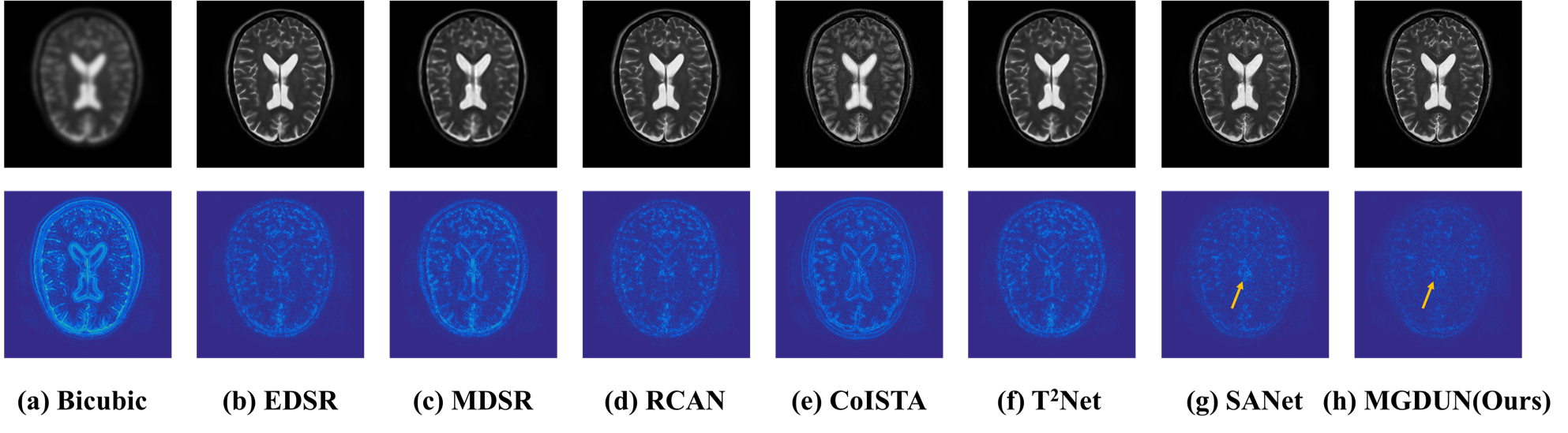}
  \caption{Qualitative comparisons of all methods on IXI dataset. The first row visualizes the visual effects of different methods, and the last row visualizes the error map between the SR results and the ground truths.}
  \label{fig:vision_ixi}
\end{figure*}

\begin{table}[]
\centering
\caption{The results of different configurations on IXI dataset. The up or down arrows indicate higher or lower values corresponding to better results.}
\label{tab:ablation}
\resizebox{\columnwidth}{!}{%
\begin{tabular}{c|c|ccc}
\hline
configuration & Stages      & PSNR $\uparrow$   & SSIM $\uparrow$  & MSE $\downarrow$   \\ \hline
I             & 3           & 36.7565 & 0.9683 & 3.7998 \\
II(Ours)      & 4           & 37.3366 & 0.9691 & 3.5598 \\
III           & 5           & 37.9190 & 0.9716 & 3.3378 \\
IV            & 6           & 38.0198 & 0.9722 & 3.3206 \\ \hline
configuration & Guide image & PSNR    & SSIM   & MSE    \\ \hline
V             & \XSolid     & 34.7081 & 0.9568 & 4.8183 \\
VI(concat)    & \XSolid     & 35.9719 & 0.9630 & 4.1703 \\
VII(Ours)     & \Checkmark  & 37.3366 & 0.9691 & 3.5598 \\ \hline
configuration & INN blocks  & PSNR    & SSIM   & MSE    \\ \hline
VIII          & 1           & 36.8480 & 0.9682 & 3.7681 \\
IX(Ours)      & 2           & 37.3366 & 0.9691 & 3.5598 \\
X             & 3           & 37.4740 & 0.9703 & 3.5038 \\ \hline
configuration & Denoiser & PSNR    & SSIM   & MSE    \\ \hline
XI(Ours)      & U-Net     & 37.3366 & 0.9691 & 3.5598 \\
XII           & Resnet    & 37.1130 & 0.9686 & 3.6551 \\ \hline
\end{tabular}%
}
\end{table}

\begin{table*}[]
\centering
\caption{The parameters and testing time results of $2\times$ enlargement}
\label{tab:params}
\resizebox{0.81\textwidth}{!}{%
\begin{tabular}{c|cccccccc}
\hline
Methods           & Bicubic & EDSR & MDSR & RCAN  & CoISTA & T$^2$Net & SANet & MGDUN\\ \hline
\# Params (M)      & -       & 1.37 & 0.34 & 12.46 & 16.43  & 0.68  & 11.41 & 1.68        \\
\# Testing time (s) & -       & 4.16 & 2.71 & 1.60  & 1.50   & 1.65  & 2.21  & 1.97        \\ \hline
\end{tabular}%
}
\end{table*}

\section{EXPERIMENT}

\subsection{Datasets}

\noindent \textbf{IXI Dataset.} The IXI dataset contains registered T2 weighted and PD weighted MR images of 578 patients. T2 weighted images were used for SR and the PD weighted images served as the guidance. We excluded a few slices of each volume as the frontal slices are much noisier than the others, making their distribution different\footnote{More details can be obtained from \url{http://brain-development.org/ixi-dataset/}.}. We splitted the IXI dataset patient-wisely into a ratio of 7:1:2 for training/validation/testing. 
The size of original HR images of both T2 and PD weighted images is $256 \times 256$.

\noindent \textbf{BraTs Dataset.} The BraTs dataset (2019) contains multimodal brain data, including registered T1, T1ce, T2, and PD weighted images. Similar to the IXI dataset, we chose T2 weighted images for SR and T1 weighted images for guidance. The size of an original HR image is $240\times240$. 3350 pairs were used for training, and 1250 paired images were used for validation. 

Finally, each T2 image was blurred with a 3×3 Gaussian filter and down-sampled. We obtained an LR image of desired dimensions after bicubic interpolation. 
Before training, for numerical stability, all images were normalized over the range of [$0, 1$].

\textbf{Metrics:} The peak signal-to-noise ratio (PSNR), structural similarity (SSIM), and mean-square error (MSE) were used to evaluate the image quality of MCSR results (the greater values indicate better performance).

\subsection{Implementation Details}

We implement MGDUN in the Pytorch framework with an NVIDIA GeForce RTX 3080TI GPU.
In the training phase, our model is trained using the ADAM optimizer with a learning rate of $1e-5$, for $200$ epochs. The parameters $\alpha$ and $\beta$ are empirically set to be 0.9 and 0.999. The batch size is set as 4. We also unfold the whole image with this approach during testing. The default stage number $T$ is set to be 4, and the number of INN blocks in the cross-modal transform module is set to 2.
And the source code is available at https://github.com/yggame/MGDUN.

\subsection{Results}
We compare our results with a number of models including classical methods, single-contrast methods, multi-contrast methods, and unfolding methods. More specifically, the proposed method is compared with Bicubic, EDSR~\cite{lim2017enhanced}, MDSR~\cite{lim2017enhanced},  RCAN~\cite{zhang2018image}, CoISTA~\cite{deng2019deep}, T$^2$Net~\cite{feng2021T2Net}, and SANet~\cite{feng2021exploring}.
The benchmark results are trained using the same ways and training datasets as described in their corresponding works.

\subsubsection{Quantitative results}
\ 
\newline

For quantitative evaluation, we utilize the average PSNR, SSIM, and MSE. 
Tab.~\ref{tab:main_metrics} reports the target contrast reconstruction performance on different datasets under $2\times$ and $4\times$ enlargements where the best and second best values are highlighted in bold red and underline blue, respectively. It is clearly noted that our method achieves the best performance on different enlargements of different datasets. The results demonstrate that our model can effectively fuse the two contrasts, which is beneficial to the SR reconstruction of the target contrast. This substantiates the effectiveness and flexibility of our method with a certain degree of generalization.

\begin{figure*}[t]
  \centering
  \includegraphics[width=0.81\linewidth]{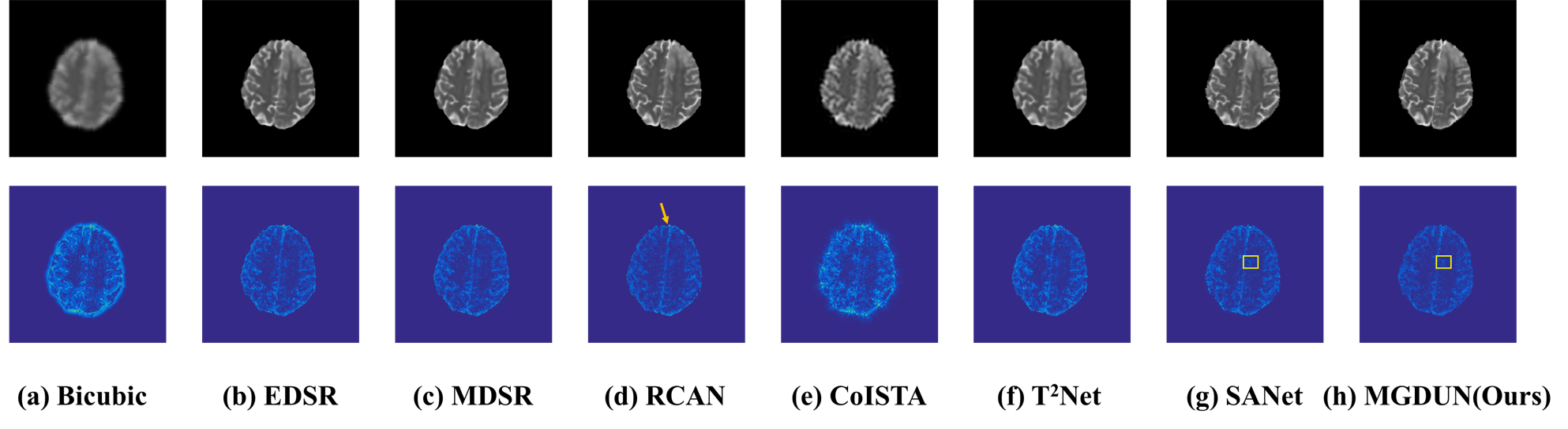}
  \caption{Qualitative comparison of all methods on BraTs dataset. The first row visualizes the visual effects of different methods, and the last row visualizes the error map between the SR results and the ground truths.}
  \label{fig:vision_brats}
\end{figure*}

\subsubsection{Qualitative results}
\ 
\newline

We provide qualitative comparison results on the IXI dataset as well as the BraTs dataset and their corresponding error maps in Fig.~\ref{fig:vision_ixi} and Fig.~\ref{fig:vision_brats}.
The texture of error maps represents the restoration error, the smoother the texture, the better the reconstruction. As we can see, the input has significant aliasing artifacts and lacks anatomical details. It can be noted that our model recovers the image with fewer visible artifacts and reconstructs more details than other competing methods. The quality improvement achieved by MDCUN may be associated with the full usage of the feature maps from the former stages to refine the final results.

\subsection{Ablation Study}

To further verify the performance of the proposed model under different configurations, a series of ablation studies are carried out, including 1) Effects of the number of stages; 2) Influence of the guide images; 3) Effect of the number of INN blocks in cross-modal transform Module; 4) Effect of the denoising modules. In this section, only IXI dataset is used.

\textbf{Effect of number of stages}~~~~
To explore the impact of the number of unfolding stages on the performance, we report the results for different realizations of the proposed model with varying number t of unfolding stages as described in Eq.~\ref{eq:iter_U} to Eq.~\ref{eq:iter_Z}.
Tab.~\ref{tab:ablation}(I-IV) shows the performance of different stages from 3 to 6. It can be observed that the performance increases as the number of stages increases. We choose $T = 4$ in our implementation to balance the performance and computational complexity.

\textbf{Influence of the guide images}~~~~
The guide images are used to provide complementary information for recovering a target modality. In order to verify the effectiveness of the guide images, we conduct a series of ablation studies (e.g., removing the guide image and using a simple multi-contrast fusion mechanism). The results are shown in Tab.~\ref{tab:ablation}(V-VII). As we can see, our approach is an effective strategy and improve performance.

\textbf{Effect of the number of INN blocks in cross-modal transform Module}~~~~
We additionally perform a comparative experiment to verify the effectiveness of INN blocks in the cross-modal transform module. As shown in Tab.~\ref{tab:ablation}(VIII-X), the performance will increase as the number of INN blocks increases.
In other words, when dealing with cross-modal transform functions, the reconstruction capability of our method could be improved by appropriately increasing the number of blocks.

\textbf{Effect of the denoising modules}
To verify the effectiveness of U-net denoising modules, we further implement an ablation study, in which U-net denoising modules are replaced by Resnet denoising modules containing a similar number of parameters. As we can see in Tab.~\ref{tab:ablation}(XI-XII), the proposed methods with U-net denoiser outperforms Resnet denoiser.

\subsection{Cost-performance Trade-off}
To demonstrate the trade-off between the cost and the performance, we compare this against several existing SCSR methods in Tab.~\ref{tab:params}. It's noted that our model can achieve better performance than others with comparable model sizes. Although it does not achieve the best performance in terms of testing time, the proposed method still has notable advantages over other competing methods. The results demonstrate that our method can yield satisfying performance with a good trade-off between cost and performance compared to other deep learning-based methods.

\section{CONCLUSION}
The interpretable deep learning model is a promising approach for the recovery of medical images as trustworthiness is required in clinical practice.
In this paper, we propose a novel Model-Guided interpretable Deep Unfolding Network (MGDUN) for medical image SR reconstruction and show how to unfold it by deep convolutional network implementation for multi-contrast medical image SR reconstruction. Our MGDUN is capable of better exploring domain knowledge in MCSR tasks and optimizing model-guided SR reconstruction algorithm in an interpretable manner.

\begin{acks}
This work was supported by the National Natural Science Foundation of China (NSFC)
under Grant 61922075, the USTC Research Funds of the Double First-Class Initiative under Grants YD2100002004 and KY2100000123 as well as the University Synergy Innovation Program of Anhui Province No. GXXT-2019-025. We acknowledge the support of GPU cluster built by MCC Lab of Information Science and Technology Institution, USTC.
\end{acks}

\newpage

\bibliographystyle{ACM-Reference-Format}
\bibliography{references}

\appendix









\end{document}